\documentclass[epr,twocolumn]{revtex4}

\usepackage{amssymb}
\usepackage{graphicx}

\begin{document}

\title{\textbf{Labyrinthine pathways towards supercycle attractors in unimodal maps}}

\author{Luis G. Moyano}
\affiliation{Departamento de Matem\'aticas and Grupo Interdisciplinar de Sistemas Complejos,
Universidad Carlos {\rm III} de Madrid, 28911 Legan\'es, Madrid, Spain}
\author{D. Silva and A. Robledo}
\affiliation{Instituto de F\'{\i}sica, Universidad Nacional Aut\'onoma de M\'exico, 
Apartado postal 20-364, M\'exico 01000 D.F., M\'exico}

\begin{abstract}
We uncover previously unknown properties of the family of periodic
superstable cycles in unimodal maps characterized each by a Lyapunov
exponent that diverges to minus infinity. Amongst the main novel properties
are the following: i) The basins of attraction for the phases of the cycles
develop fractal boundaries of increasing complexity as the period-doubling
structure advances towards the transition to chaos. ii) The fractal
boundaries, formed by the preimages of the repellor, display hierarchical
structures organized according to exponential clusterings that manifest in
the dynamics as sensitivity to the final state and transient chaos. iii)
There is a functional composition renormalization group (RG) fixed-point map
associated to the family of supercycles. iv) This map is given in closed
form by the same kind of $q$-exponential function found for both the
pitchfork and tangent bifurcation attractors. v) There is a final stage
ultra-fast dynamics towards the attractor with a sensitivity to initial
conditions that decreases as an exponential of an exponential of time.

Key words: supercycles, fractal boundaries, transient chaos, RG fixed-point
map

PACS: 05.45.-a, 64.60.Ht, 05.45.Df, 02.60.Cb
\end{abstract}
\maketitle
\section{Introduction}

The family of superstable orbits - often called supercycles - of unimodal
maps have played a prominent role in the historical uncovering of the
cascade of period-doubling bifurcations that leads to the transition to
chaos with the celebrated universal features displayed by the Feigenbaum
attractor \cite{schuster1}, \cite{beck1}, \cite{hilborn1}. These periodic
attractors are easy to find because one of its cycle positions corresponds
to the maximum of the map. And then, they are convenient for basic
computational purposes because trajectories originating at different
positions settle very fast into the attractor for the reason that their
Lyapunov exponent diverges to minus infinity. And so they offered a suitable
path for the initial characterization of the period-doubling route to chaos.
Oddly enough, several features associated to the dynamical evolution toward
these attractors have remained unexplored. Here we provide an account of
these properties, some shared by other families of periodic attractors and
some specific of the supercycles. We discuss explicitly the case of a map
with quadratic maximum but the results are easily extended to general
nonlinearity $z>1$. In closing we indicate the broader theoretical interest
in these properties.

In order to obtain dynamical properties with previously unstated detail we
determined the organization of the \textit{entire} set of trajectories as
generated by all possible initial conditions. We find that the paths taken
by the full set of trajectories in their way to the supercycle attractors
(or to their complementary repellors) are far from unstructured. The
preimages of the attractor of period $2^{N}$, $N=1,2,3, \ldots$ are distributed
into different basins of attraction, one for each of the $2^{N}$\ phases
(positions) that compose the\ cycle. When $N\geq 2$ these basins are
separated by fractal boundaries whose complexity increases with increasing $N
$. The boundaries consist of the preimages of the corresponding repellor and
their positions cluster around the $2^{N}-1$ repellor positions according to
an exponential law. As $N$ increases the structure of the basin boundaries
becomes more involved. Namely, the boundaries for the $2^{N}$ cycle develops
new features around those of the previous $2^{N-1}$ cycle boundaries, with
the outcome that a hierarchical structure arises, leading to embedded
clusters of clusters of boundary positions, and so forth.

The dynamics associated to families of trajectories always displays a
distinctively concerted order that reflects the repellor preimage boundary
structure of the basins of attraction. That is, each trajectory has an
initial condition that is identified as an attractor (or repellor) preimage
of a given order, and this trajectory necessarily follows the steps of other
trajectories with initial conditions of lower preimage order belonging to a
given chain or pathway to the attractor (or repellor). This feature gives
rise to transient chaotic behavior different from that observed at the last
stage of approach to the attractor. When the period $2^{N}$ of the cycle
increases the dynamics becomes more involved with increasingly more complex
stages that reflect the preimage hierarchical structure. As a final point,
at the closing part of the last leg of the trajectories an ultra-rapid
convergence to the attractor is observed, with a sensitivity to initial
conditions that decreases as an exponential of an exponential of time. In
relation to this we find that there is a functional composition
renormalization group (RG) fixed-point map associated to the supercycle
attractor, and this can be expressed in closed form by the same kind of $q$
-exponential function found for both the pitchfork and tangent bifurcation
attractors \cite{robledo1}, \cite{robledo2}, like that originally derived by
Hu and Rudnick for the latter case \cite{hu1}.

\begin{figure}[h!]
\centering
\includegraphics[width=.5\textwidth]{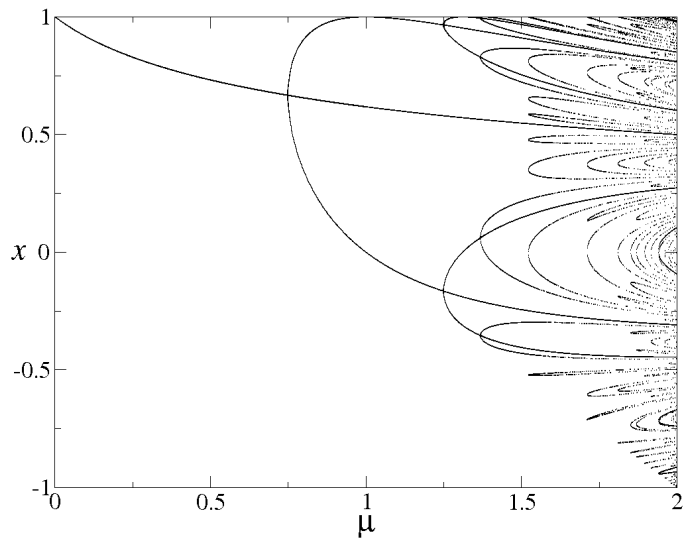}
\caption{
{Solutions of }${\small x=f}_{\mu }^{(2^{N})}{\small (x)}$
{\small , }${\small N=0,1,2,...}$, {\small where }${\small f}_{\mu }${\small 
\ is the logistic map, as a function of the control parameter }${\small \mu }
${\small . The attractors (stable solutions) become unstable at each
bifurcation, turning into repellors. When for a given value of }${\small \mu 
}${\small \ the attractor has }${\small 2}^{N}${\small \ positions there are 
}${\small 2}^{N}{\small -1}${\small \ repellor positions.}
}
\label{fig1}
\end{figure}

Before proceeding to give details in the following sections of the
aforementioned dynamics we recall basic features of the bifurcation forks
that form the period-doubling cascade sequence in unimodal maps, epitomized
by the logistic map $f_{\mu }(x)=1-\mu x^{2}$, $-1\leq x\leq 1$, $0\leq \mu
\leq 2$ \cite{schuster1}, \cite{beck1}, \cite{hilborn1}. The superstable
periodic orbits of lengths $2^{N}$, $N=1,2,3, \ldots$, are located along the
bifurcation forks, i.e. the control parameter value $\mu =\overline{\mu }%
_{N}<\mu _{\infty}$ for the superstable $2^{N}$-attractor is that for which
the orbit of period $2^{N}$ contains the point $x=0$, where $\mu _{\infty
}=1.401155189\ldots$ is the value of $\mu $ at the period-doubling accumulation
point. The positions (or phases) of the $2^{N}$-attractor are given by $%
x_{m}=f_{\overline{\mu }_{N}}^{(m)}(0)$, $m=1,2, \ldots,2^{N}$. Notice that
infinitely many other sequences of superstable attractors appear at the
period-doubling cascades within the windows of periodic attractors for
values of $\mu >$ $\mu _{\infty }$. Associated to the $2^{N}$-attractor at $%
\mu =\overline{\mu }_{N}$ there is a $(2^{N}-1)$-repellor consisting of $%
2^{N}-1$ positions $y_{m}$, $m=1,2, \ldots,2^{N}-1$. These positions are the
unstable solutions, $\left\vert df_{\overline{\mu }_{N}}^{(2^{n-1})}(y)/dy%
\right\vert <1$ , of $y=f_{\overline{\mu }_{N}}^{(2^{n-1})}(y)$, $%
n=1,2, \ldots,N $. The first, $n=1$, originates at the initial period-doubling
bifurcation, the next two, $n=2$, start at the second bifurcation, and so
on, with the last group of $2^{N-1}$, $n=N$, stemming from the $N-1$
bifurcation. See Fig. \ref{fig1}. We find it useful to order the repellor positions,
or simply, repellors, present at $\mu =\overline{\mu }_{N}$, according to a
hierarchy or tree, the `oldest' with $n=1$ up to the most recent ones with $%
n=N$. The repellors' order is given by the value of $n$. Finally, we define
the preimage $x^{(k)}$ of order $k$ of position $x$ to satisfy $%
x=h^{(k)}(x^{(k)})$ where $h^{(k)}(x)$ is the $k$-th composition of the map $%
h(x)\equiv f_{\overline{\mu }_{N}}^{(2^{N-1})}(x)$. We have omitted
reference to the $2^{N}$-cycle in $x^{(k)}$ to simplify the notation.

\section{Preimage structure of supercycle attractors}

\begin{figure}[h!]
\centering
\includegraphics[width=.5\textwidth]{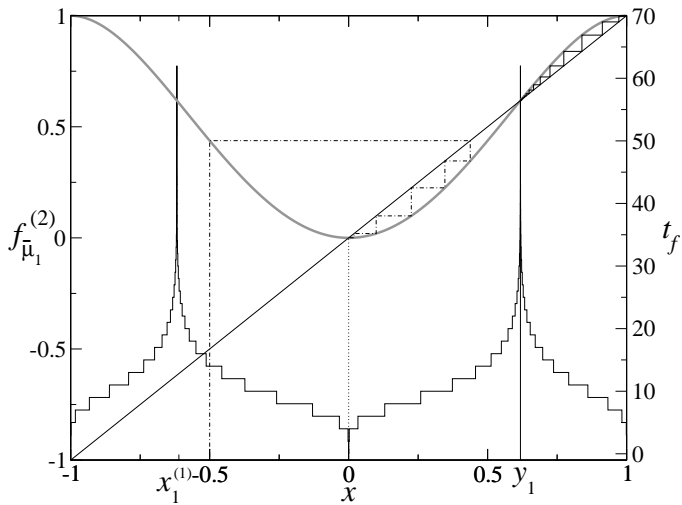}
\caption{
{\small Left axis: The twice iterated map }${\small f}_{\overline{%
\mu }_{1}}^{(2)}{\small (x)}$, $\overline{\mu }_{1}{\small =1}${\small \ (in
grey line)}.{\small \ Right axis: Time of flight }${\small t}_{f}{\small (x)}
${\small , the number of iterations necessary for a trajectory with initial
condition at }${\small x}${\small \ to reach an attractor position. The
values of }${\small x}${\small \ near the high spikes correspond to initial
conditions very close to the repellor and its preimage. We present three
example trajectories (and the }${\small y=x}${\small \ line as an aid to
visualize them): The dotted line shows a trajectory that starts at the
attractor position }${\small x=0}${\small \ and remains there. The solid
line is a trajectory starting near the repellor at }${\small y}_{{\small 1}}$%
{\small , and after a large number of iterations reaches the attractor
position }${\small x=1}${\small . Finally, the dash-dotted line is an orbit
starting at }${\small x=0.5}${\small \ that in just a few iterations reaches 
}${\small x=0}${\small .}
}
\label{fig2}
\end{figure}

The core source of our description of the dynamics towards the supercycle
attractors is a measure of the relative `time of flight' $t_{f}(x_{0})$ for
a trajectory with initial condition $x_{0}$ to reach the attractor. The
function $t_{f}(x_{0})$ is obtained for an ensemble representative of 
\textit{all} initial conditions $-1\leq x_{0}\leq 1$. This comprehensive
information is determined through the numerical realization of every
trajectory, up to a small cutoff $\varepsilon >0$ at its final stage. The
cutoff $\varepsilon $ regards a position $x_{f}\leq x_{m}\pm \varepsilon $
to be effectively the attractor phase $x_{m}$. This, of course, introduces
an approximation to the real time of flight, that can be arbitrarily large
for those $x_{0}$ close to a repellor position $y_{m}$ or close to any of
its infinitely many preimages, $x_{m}^{(k)}$, $k=1,2, \ldots$, $N\geq 2$. In
such cases the finite time $t_{f}(x_{0};\varepsilon )$ can be seen to
diverge $t_{f}\rightarrow \infty $ as $x_{0}\rightarrow y_{m}$ and $%
\varepsilon \rightarrow 0$. As a simple illustration, in Fig. \ref{fig2} we show the
time of flight $t_{f}(x_{0})$ for the period $2$ supercycle at $\mu =%
\overline{\mu }_{1}$ with $\varepsilon =10^{-9}$ together with the
twice-composed map $f_{\overline{\mu }_{1}}^{(2)}(x)$ and a few representative
trajectories. We observe two peaks in $t_{f}(x_{0};\varepsilon )$ at $%
y_{1}=(-1+\sqrt{1+4\overline{\mu }_{1}})/2\overline{\mu }_{1} \simeq 0.6180340\ldots$, 
the fixed-point repellor and at its (only) preimage $
x_{1}^{(1)}=-y_{1} $, where $y_{1}=f_{\overline{\mu }_{1}}^{(2)}(x_{1}^{(1)})$.
Clearly, there are two basins of attraction each for the two positions or
phases, $x_{1}=0$ and $x_{2}=1$, of the attractor. For the former it is the
interval $x_{1}^{(1)}<x_{0}<y_{1}$, whereas for the second one consists of
the intervals $-1\leq x_{0}<x_{1}^{(1)}$ and $y_{1}<x_{0}\leq 1$. The
multiple-step structure of $t_{f}(x_{0})$, of four time units each step,
reflects the occurrence of intervals of initial conditions with common
attractor phase preimage order $k$.

\begin{figure}[h!]
\centering
\includegraphics[width=.5\textwidth]{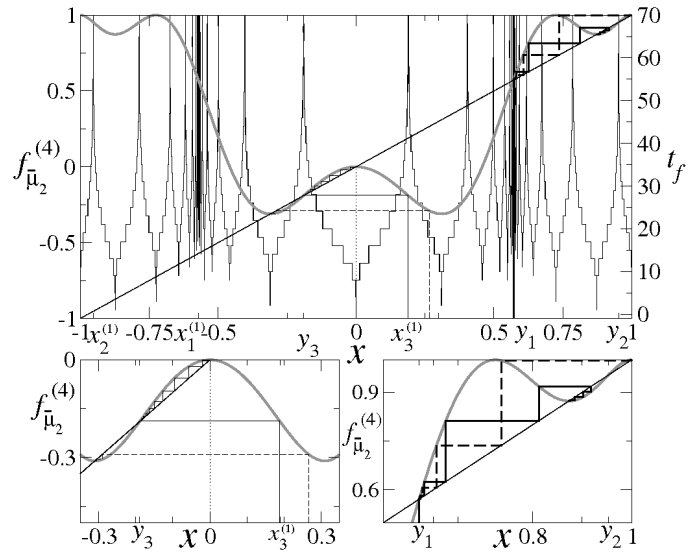}
\caption{
{\small Same as Fig. \ref{fig2} but for }$f_{\overline{\mu }_{2}}^{(4)}(x)$, $%
\overline{\mu }_{2}\simeq {\small 1.31070264}${\small . Two orbits (in solid
bold and dashed bold lines) start very near each other and by the position }$%
{\small y}_{{\small 1}}${\small \ of the old repellor, indistinguishable at
the beginning they take very different subsequent paths to reach the
attractor positions at }${\small x=1}${\small \ and }${\small x\simeq 0.8734}
${\small . See bottom right panel. Another two orbits (in dotted and solid
lines) start at an attractor position }${\small x=1}${\small \ and at a
repellor preimage positions, respectively. Finally, one more orbit (in
dashed line) starts at an intermediate initial condition an reaches very
quickly the attractor position at }${\small x\simeq -0.310703}${\small . See
bottom left panel.}
}
\label{fig3}
\end{figure}

\begin{figure}[h!]
\centering
\includegraphics[width=.5\textwidth]{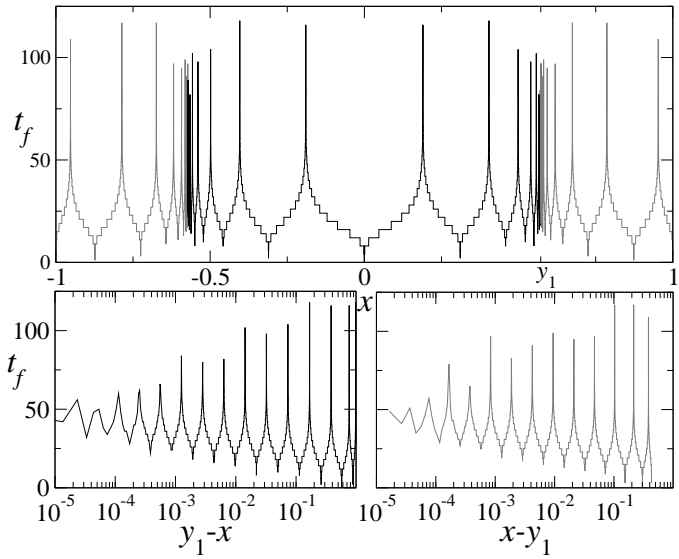}
\caption{
{\small Top panel: Time of flight }${\small t}_{f}{\small (x)}$%
{\small \ for }${\small N=2}${\small \ (as in Fig. 3), the black lines
correspond to initial conditions that terminate at the attractor positions }$%
{\small x=0}${\small \ and }${\small x\simeq -0.310703}${\small , while the
grey lines to trajectories ending at }${\small x=1}${\small \ and }${\small %
x\simeq 0.8734}${\small . Right (left) bottom panel: Same as top panel, but
plotted against the logarithm of }${\small x-y}_{{\small 1}}${\small \ (}$%
{\small y}_{{\small 1}}{\small -x}${\small ). It is evident that the peaks
are arranged exponentially around the old repellor position }${\small y}_{%
{\small 1}}${\small , i.e., they appear equidistant in a logarithmic scale.
See Fig. \ref{fig5}.}
}
\label{fig4}
\end{figure}

\begin{figure}[h!]
\centering
\includegraphics[width=.5\textwidth]{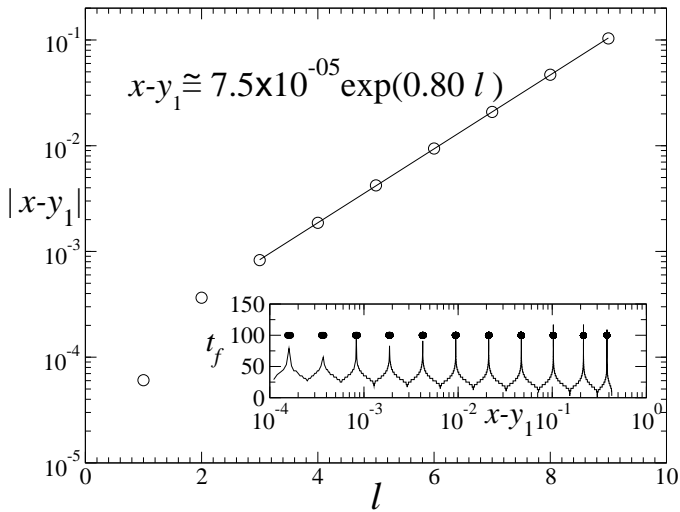}
\caption{
{\small Corroboration of preimage exponential clustering around the
repellor position }${\small y}_{{\small 1}}\simeq {\small 0.571663}${\small %
\ when }${\small N=2}${\small \ and }$\overline{\mu }_{2}\simeq {\small %
1.31070264}$.{\small \ The variable }${\small l}${\small \ labels
consecutive equidistant peaks in the inset. The peaks correspond to
preimages of the repellor. See text.}
}
\label{fig5}
\end{figure}

For the next supercycle - period $4$ - the preimage structure turns out to
be a good deal more involved than the straightforward structure for $%
\overline{\mu }_{1}$. In Fig. \ref{fig3} we show the times of flight $t_{f}(x_{0})$
for the $N=2$ supercycle at $\mu =\overline{\mu }_{2}$ with $\varepsilon
=10^{-9}$, the map $f_{\overline{\mu }_{2}}^{(4)}(x)$ is superposed as a
reference to indicate the four phases of the attractor (at $x_{1}=0$, $%
x_{2}=1$, $x_{3}\simeq -0.3107\ldots$, and $x_{4}\simeq 0.8734\ldots$) and the three
repellor positions (at $y_{1}\simeq 0.5716635\ldots$, $y_{2}\simeq 0.952771\ldots$, and $
y_{3}\simeq -0.189822\ldots$). In Fig. \ref{fig3} there are also shown four trajectories each of
which terminates at a different attractor phase. We observe a proliferation
of peaks and valleys in $t_{f}(x_{0})$, actually, an infinite number of
them, that cluster around the repellor at $y_{1}\simeq 0.5716635\ldots$ and also at its
preimage at $x_{1}^{(1)}=-y_{1}$ (these are the positions at $\mu =
\overline{\mu }_{2}$ of the `old' repellor and its preimage in the previous $
N=1$ case). Notice that the steps in the valleys of $t_{f}(x_{0})$ are now
eight time units each. The nature of the clustering of peaks (repellor phase
preimages) and the bases of the valleys (attractor phase preimages) is
revealed in Fig. \ref{fig4} where we plot $t_{f}$ in a logarithmic scale for the
variables $\pm (x-y_{1})$. There is an exponential clustering of the
preimage structure around both the old repellor and its preimage. This
scaling property is corroborated in Fig. \ref{fig5} from where we obtain $%
x-y_{1}\simeq 7.5\times10^{-5}\exp (0.80\,l)$, where $l=1,2,3, \ldots$ is a label for
consecutive repellor preimages.

\begin{figure}[h!]
\centering
\includegraphics[width=.5\textwidth]{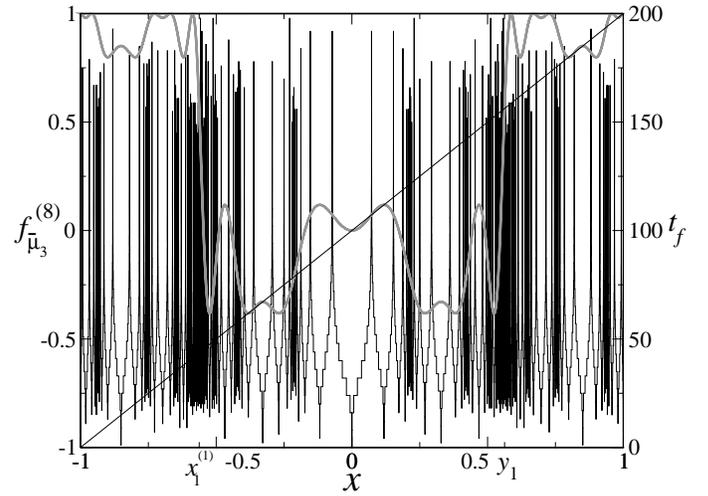}
\caption{
{\small Same as Fig. \ref{fig3} for }${\small f}_{\overline{\mu }_{3}}^{(8)}%
{\small (x)}$, $\overline{\mu }_{3}{\small \simeq 1.38154748}$.{\small \
Here }${\small y}_{{\small 1}}{\small =0.56264475}${\small .}
}
\label{fig6}
\end{figure}

\begin{figure}[h!]
\centering
\includegraphics[width=.5\textwidth]{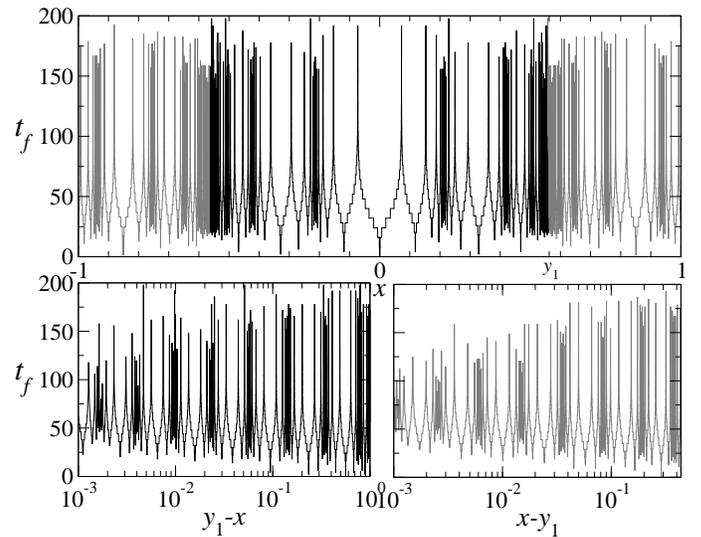}
\caption{
{\small Same as Fig. \ref{fig4} for }${\small N=3}${\small . The black lines
correspond to initial conditions that terminate at any of the four attractor
positions close or equal to }${\small x=0}${\small , while the grey lines to
trajectories ending at any of the other four attractor positions close or
equal to }${\small x=1}$. {\small As the bottom panels show, in logarithmic
scale, in this case there are (infinitely) many clusters of peaks (repellor
preimages) equidistant from each other.}
}
\label{fig7}
\end{figure}

\begin{figure}[h!]
\centering
\includegraphics[width=.5\textwidth]{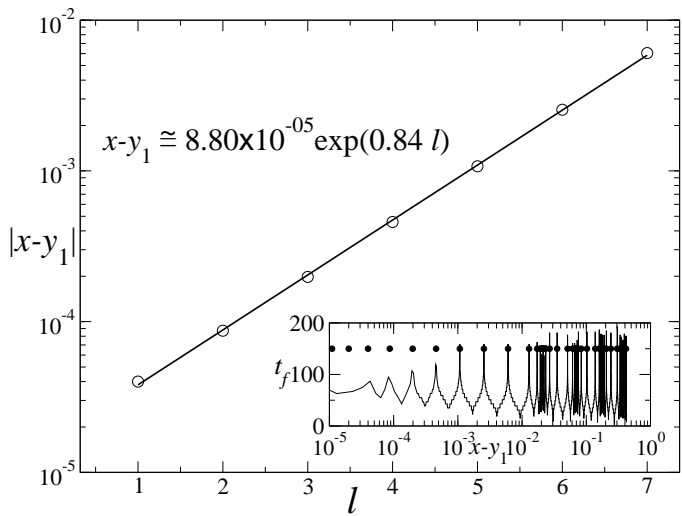}
\caption{
{\small Same as Fig. \ref{fig5} for }${\small N=3}${\small .}
}
\label{fig8}
\end{figure}

A comparable leap in the complexity of the preimage structure is observed
for the following - period $8$ - supercycle. In Fig. \ref{fig6} we show $t_{f}(x_{0})$
for the $N=3$ supercycle at $\mu =\overline{\mu }_{3}$ with $\varepsilon
=10^{-9}$, together with the map $f_{\overline{\mu }_{2}}^{(8)}(x)$ placed
as reference to facilitate the identification of the locations of the eight
phases of the attractor, $x_{1}$ to $x_{8}$, and the seven repellor
positions, $y_{1}$ to $y_{7}$. Besides a huge proliferation of peaks and
valleys in $t_{f}(x_{0})$, we observe now the development of clusters of
clusters of peaks centered around the repellor at $y_{1}\simeq 0.5626447\ldots$ and its
preimage at $x_{1}^{(1)}=-y_{1}$ (these are now the positions of the
original repellor and its preimage for the $N=1$ case when $\mu =\overline{%
\mu }_{3}$). The steps in the valleys of $t_{f}(x_{0})$ have become sixteen
time units each. Similarly to the clustering of peaks (repellor phase
preimages) and valleys (attractor phase preimages) for the previous
supercycle at $\overline{\mu }_{2}$, the spacing arrangement of the new
clusters of clusters of peaks is determined in Fig. \ref{fig7} where we plot $t_{f}$
in a logarithmic scale for the variables $\pm (x-y_{1})$. In parallel to the
previous cycle an exponential clustering of clusters of the preimage
structure is found around both the old repellor and its preimage. This
scaling property is quantified in Fig. \ref{fig8} from where we obtain $x-y_{1}\simeq
8.8\times 10^{-5}\exp (0.84\,l),$ where $l=1,2,3, \ldots$ counts consecutive clusters.

\begin{figure}[h!]
\centering
\includegraphics[width=.5\textwidth]{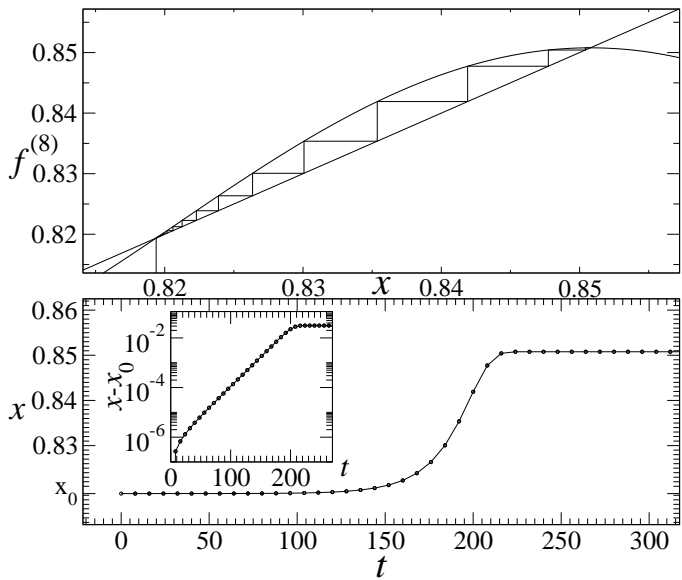}
\caption{
{\small Top panel: Detail of the map }${\small f}_{\overline{\mu }%
_{3}}^{(8)}{\small (x)}${\small \ and a trajectory in the proximity of the
attractor position located at }${\small x\simeq 0.850780}$. {\small The
trajectory originated very close to the repellor position at }${\small %
x=0.819378}${\small . Bottom panel: The same trajectory of the
eight-iterated map as a function of time }${\small t}${\small . Inset:
Corroboration of the exponential nature of the trajectory after it leaves
the repellor and before the final approach to the attractor position.}
}
\label{fig9}
\end{figure}

An investigation of the preimage structure for the next $N=4$ supercycle at $%
\mu =\overline{\mu }_{4}$ leads to another substantial increment in the
involvedness of the structure of preimages but with such density that makes
it cumbersome to describe here. Nevertheless it is clear that the main
characteristic in the dynamics is the development of a hierarchical
organization of the preimage structure as the period $2^{N}$ of the
supercycles increases.

\section{Final state sensitivity and transient chaos}

With the knowledge gained about the features displayed by the times of
flight $t_{f}(x_{0})$ for the first few supercycles it is possible to
determine and understand how the leading properties in the dynamics of
approach to these attractors arise. The information contained in $%
t_{f}(x_{0})$ can be used to demonstrate in detail how the concepts of final
state sensitivity \cite{grebogi1} - due to attractor multiplicity - and
transient chaos \cite{beck1} - prevalent in the presence of repellors that
coexist with periodic attractors - realize in a given dynamics. Final state
sensitivity is the consequence of fractal boundaries separating coexisting
attractors. In our case there is always a single attractor but its positions
or phases play an equivalent role \cite{grebogi2}. Transient chaos \cite%
{beck1} is due to fast separation in time of nearby trajectories by the
action of a repellor and results in a sensitivity to initial conditions that
grows exponentially up to a crossover time after which the decay sets in. We
describe below how both properties result from an extremely ordered flow of
trajectories towards the attractor. This order is imprinted by the preimage
structure described in the previous section.

For the simplest supercycle at $\mu =\overline{\mu }_{1}$ there is trivial
final state sensitivity as the boundary between the two basins of the
phases, $x_{1}=0$ and $x_{2}=1$, consists only of the two positions, $%
y_{1}\simeq 0.5716635\ldots$ and $x_{1}^{(1)}=-y_{1}$ See Fig. \ref{fig2}. Consider the
length $\delta $ of a small interval around a given value of $x_{0}$
containing either $y_{1}$ or $x_{1}^{(1)}$, when $\delta \rightarrow 0$ any
uncertainty as to the final phase of the trajectory disappears. It is also
simple to verify that when $x_{0}$ is close to $y_{1}$ or $x_{1}^{(1)}$ the
resulting trajectories increase their separation at initial and intermediate
times displaying transient chaos in a straightforward fashion. In Fig. \ref{fig9} we
show the same type of transitory exponential sensitivity to initial
conditions for a trajectory at $\mu =\overline{\mu }_{3}$ after it reaches a
repellor position in the final journey towards the period $8$ attractor.
This behavior is common to all periodic attractors when trajectories come
near a repellor at the final leg of their journey.

\begin{figure}[h!]
\centering
\includegraphics[width=.5\textwidth]{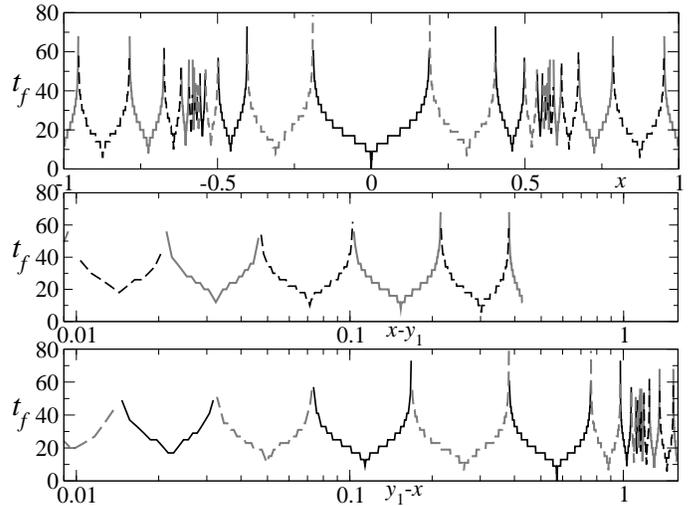}
\caption{
{\small Time of flight }${\small t}_{f}{\small (x)}${\small \ for }$%
{\small N=2}$ {\small (equivalent to Fig. \ref{fig4}). Different types of lines are
used for different sub-basins of attraction that form the fractal boundary
between the attractor positions. The black solid line corresponds to the
attractor position }${\small x=0}${\small , the grey solid line to }${\small %
x=1}${\small , dashed grey line to }${\small x\simeq -0.310703}${\small ,
and dashed black line to }${\small x\simeq 0.873470}${\small . The
sub-basins are separated in an alternated fashion by repulsor preimages
(peaks) and there are just two types of sub-basins depending on }${\small |x|%
}${\small \ being larger or smaller than }${\small y}_{{\small 1}}${\small .}
}
\label{fig10}
\end{figure}

For $\mu =\overline{\mu }_{2}$ there are more remarkable properties arising
from the more complex preimage structure. There is a concerted migration of
initial conditions seeping through the boundaries between the four basins of
attraction of the phases. These boundaries, shown in Fig. \ref{fig10}, form a fractal
network of interlaced sub-basins separated from each other by two preimages
of different repellor phases and have at their bottom a preimage of an
attractor phase. Trajectories on one of these sub-basins move to the nearest
sub-basin of its type (next-nearest neighbor in actual distance in Fig. \ref{fig10})
at each iteration of the map $f_{\overline{\mu }_{2}}^{(4)}$ ($4$ time steps
for the original map $f_{\overline{\mu }_{2}}$). The movement is always away
from the center of the cluster at the old repellor position $y_{1}$ or at
its preimage $x_{1}^{(1)}$ (located at the steepest slope inflection points
of $f_{\overline{\mu }_{2}}^{(4)}$ shown in Fig. \ref{fig3}). Once a trajectory is
out of the cluster (contained between the maxima and minima of $f_{\overline{%
\mu }_{2}}^{(4)}$ next to the mentioned inflection points) it proceeds to
the basin of attraction of an attractor phase (separated from the cluster by
the inflection points with gentler slope of $f_{\overline{\mu }_{2}}^{(4)}$
in Fig. \ref{fig3}) where its final stage takes place. When considering a large
ensemble of initial positions, distributed uniformly along all phase space,
the common journey towards the attractor displays an exceedingly ordered
pattern. Each initial position $x_{0}$ within either of the two clusters of
sub-basins is a preimage of a given order $k$ of a position in the main
basin of attraction. Each iteration of $f_{\overline{\mu }_{2}}^{(4)}$
reduces the order of the preimage from $k$ to $k-1$, and the new position $%
x_{0}^{\prime }=f_{\overline{\mu }_{2}}^{(4)}(x_{0})$ replaces the initial
position $x_{0}^{\prime }$ (a preimage of order $k-1$) of another trajectory
that under the same time step has migrated to the initial position $%
x_{0}^{\prime \prime }$ (a preimage of order $k-2$) of another trajectory,
and so on.

\begin{figure}[h!]
\centering
\includegraphics[width=.5\textwidth]{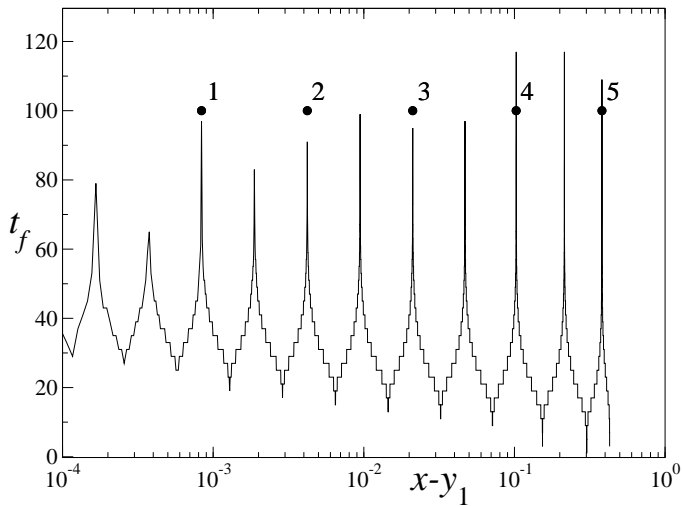}
\caption{
{\small Evolution of an orbit starting very close to a peak
(repellor preimage) inside a cluster when }${\small N=2}${\small , }$%
\overline{\mu }_{2}\simeq {\small 1.31070264}${\small \ and }${\small y}_{1}%
{\small =0.571663}${\small . (See Fig. \ref{fig4}).The trajectory describes an
exponential in time moving away from }${\small y}_{{\small 1}}${\small . In
the figure, positions of the trajectory correspond to the black solid dots,
and iterations correspond to the number associated to each dot.}
}
\label{fig11}
\end{figure}

It is clear that the dynamics at $\mu =\overline{\mu }_{2}$ displays
sensitivity to the final state when the initial condition $x_{0}$ is located
near the core of any of the two clusters of sub-basins at $y_{1}$ and $%
x_{1}^{(1)}$ that form the boundary of the attractor phases. Any uncertainty
on the location of $x_{0}$ when arbitrarily close to these positions implies
uncertainty about the final phase of a trajectory. There is also transient
chaotic behavior associated to the migration of trajectories out of the
cluster as a result of the organized preimage resettlement mentioned above.
Indeed, the exponential disposition of repellor preimages shown in Fig. \ref{fig5} is
actually a realization of two trajectories with initial conditions in
consecutive peaks of the cluster structure. Therefore, the exponential
expression given in the previous Section in relation to Fig. \ref{fig5} can be
rewritten as the expression for a trajectory, $x_{\tau }\simeq x_{0}\exp
(\lambda _{eff}\tau )$, with, $x_{\tau }=x-y_{1}$, $x_{0}=$ $7.5\times 10^{-5}$,
and $\lambda _{eff}=6.4$, where $\tau =1,2,3,\ldots$. Straightforward
differentiation of $x_{\tau }$ with respect to $x_{0}$ yields an exponential
sensitivity to initial conditions with positive effective Lyapunov
coefficient $\lambda _{eff}$.

\begin{figure}[h!]
\centering
\includegraphics[width=.5\textwidth]{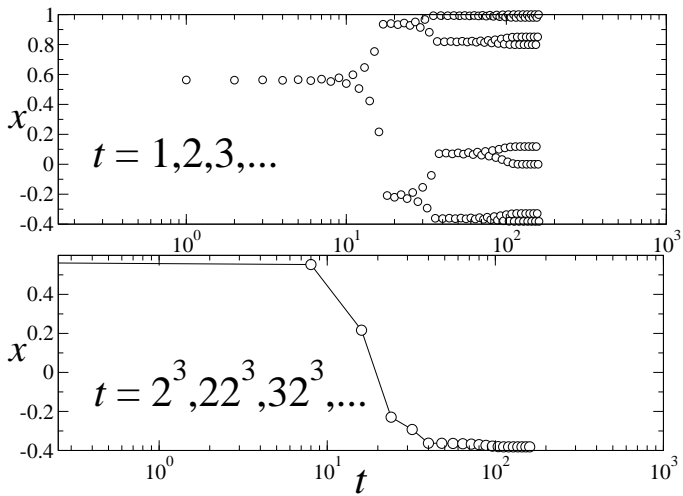}
\caption{
{\small A trajectory for }$\overline{\mu }_{3}{\small \simeq
1.38154748}${\small . Top panel: The circles are positions for consecutive
times in the iterations of the map }${\small f}_{\overline{\mu }_{3}}{\small %
(x)}${\small . The orbit starts very close to the old (period one) repellor }%
${\small y}_{{\small 1}}${\small , then moves close to a period two and
subsecuently to a period four repellor, until finally the trajectory arrives
at the period eight attractor. Bottom panel: Selection of a time subsequence
of multiples of }${\small 2}^{3}${\small \ shows the same trajectory as an
evolution towards one particular attractor position.}
}
\label{fig12}
\end{figure}

As it can be anticipated, the dynamics of approach to the next supercycle at 
$\mu =\overline{\mu }_{3}$ can be explained by enlarging the description
presented above for $\mu =\overline{\mu }_{2}$ with the additional features
of its preimage structure already detailed in the previous section. As in
the previous case, trajectories with initial conditions $x_{0}$ located
inside a cluster of sub-basins of the attractor phases will proceed to move
out of it in the systematic manner described for the only two isolated
clusters present when $\mu =\overline{\mu }_{2}$. Only now there is an
infinite number of such clusters arranged into two bunches that group
exponentially around the old repellor position $y_{1}$ and around its
preimage $x_{1}^{(1)}$. See Figs. \ref{fig6} to \ref{fig8}. Once such trajectories leave the
cluster under consideration they enter into a neighboring cluster, and so
forth, so that the trajectories advance out of these fractal boundaries
through the prolonged process of migration out of the cluster of clusters
before they proceed to the basins of the attraction of the eight phases of
this cycle. In Fig. \ref{fig12} we show one such trajectory in consecutive times $%
t=1,2,3, \ldots$ for the original map and also in multiples of time $%
t=2^{3},22^{3},32^{3}, \ldots$ The logarithmic scale of the figure makes evident
the retardation of each stage in the process. As when $\mu =\overline{\mu }%
_{2}$, it is clear that in the approach to the $\mu =\overline{\mu }_{3}$
attractor there is sensitivity to the final state and transitory chaotic
sensitivity to initial conditions. Again, the exponential expression, given
in the previous Section associated to the preimage structure of clusters of
clusters of sub-basins shown in Fig. \ref{fig7}, can be interpreted as the expression
of a trajectory of the form $x_{\tau }\simeq x_{0}\exp (\lambda _{eff}\tau )$%
. Differentiation of $x_{\tau }$ with respect to $x_{0}$ yields again an
exponential sensitivity to initial conditions with positive effective
Lyapunov coefficient $\lambda _{eff}$.

When the period $2^{N}$ of the cycles increases we observe that the main
characteristic in the dynamics is the development of a hierarchical
organization in the flow of an ensemble of trajectories out of an
increasingly more complex disposition of the preimages of the attractor
phases. The role played by a cluster of sub-basins when $\mu =\overline{\mu }%
_{2}$ is preserved for the clusters present when $\mu =\overline{\mu }_{3}$
but the additional presence of clusters of clusters of sub-basins introduces
a similar role in the dynamics at a broader scale in the preimage structure.
A complication that will arise in a comparable fashion every time $N$
increases.

\section{Super strong insensitivity to initial conditions}

We turn now to discuss the closing point of the approach to the supercycle
attractors and find out what is the form of the sensitivity to initial
conditions $\xi _{t}$ in the $t\gg1$ limit, as required in its usual
definition. The question is pertinent because for these attractors the
ordinary Lyapunov exponent $\lambda \equiv \lim_{t\rightarrow \infty
}t^{-1}\ln \left. df_{\overline{\mu }_{N}}^{(t)}(x)/dx\right\vert _{x=0}$
diverges to minus infinity ($df_{\overline{\mu }_{N}}^{(t)}(x)/dx=0$ at $x=0$%
) and $\xi _{t}$ cannot have an exponential form $\xi _{t}=\exp (\lambda t)$
with $\lambda <0$.

\begin{figure}[h!]
\centering
\includegraphics[width=.5\textwidth]{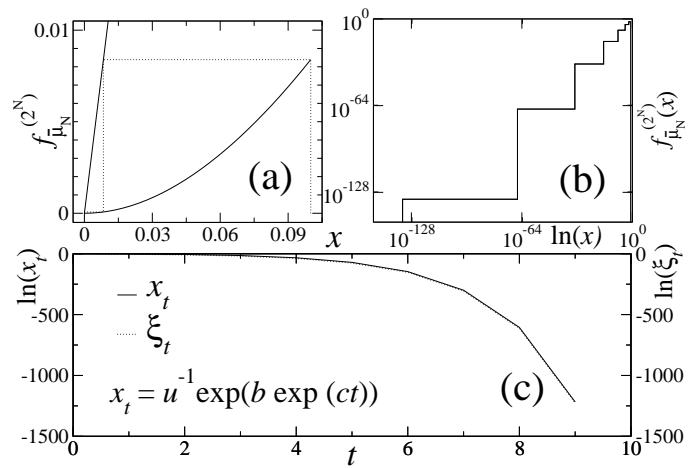}
\caption{
{\small Panel (a): Detail of a trajectory in its final stage of
approach to the }${\small x=0}${\small \ attractor position (dotted line)
for the map }$f_{\overline{\mu }_{N}}^{(2^{N})}(x)${\small \ (solid line).
In this example }${\small N=3}${\small . Panel (b): Same as (a) in double
logarithmic scale. Panel (c): Trajectory }$x_{t}${\small \ and sensitivity
to initial conditions }$\xi _{t}${\small \ in logarithmic scale versus time }%
${\small t}${\small . Both functions are indistinguishable. Note that there
is an ultra-rapid convergence, of only a few time steps, to the origin }$%
{\small x=0}${\small . See text.}
}
\label{fig13}
\end{figure}

Representative results for the last segment of a trajectory and the
corresponding sensitivity $\xi _{t}$ obtained from a numerical investigation
are shown in Fig. \ref{fig13}.\ Only the first two steps of a trajectory with $%
x_{0}=0.1$ of the map $f_{\overline{\mu }_{3}}^{(2^{3})}$ can be seen in
Fig. \ref{fig13} (a). A considerable enlargement of the spatial scale (that requires
computations of extreme precision [9]) makes it possible to observe a total
of seven steps ($56$ iterations in the original map), as shown with the help
of logarithmic scales in Fig. \ref{fig13} (b). Fig. \ref{fig13} (c) shows both 
the same trajectory
and the sensitivity $\xi _{t}\equiv dx_{t}/dx_{0}$ in a logarithmic scale
for $x_{t}$ and $\xi_{t}$ and a normal scale for the time $t$. The
trajectory is accurately reproduced (indistinguishable from the curve in
Fig. \ref{fig13} (c)) by the expression $x_{t}=u^{-1}\exp (b\exp ct)$, $b=\ln ux_{0}$
(with $x_{0}>0$) and $c=\ln 2$, where $u>0$ is obtained from the form $|f_{%
\overline{\mu }_{N}}^{(2^{N})}|\simeq ux^{2}$ that the $2^{N}$-th composed
map takes close to $x=0$. This expression for\ $x_{t}$ is just another form
of writing $ux_{t}=(ux_{0})^{2^{t}}$, the result of repeated iteration of $%
ux^{2}$. For the logarithm of the sensitivity we have $\ln \xi _{t}=-\ln
x_{0}+t\ln 2+\ln x_{t}$ where the last (large negative) term dominates the
first two. Thus, we find that the sensitivity decreases faster than an
exponential, and more precisely, decreases as the exponential of an
exponential.

It is interesting to note that, associated to the general form $f_{\overline{\mu }%
_{N}}^{(2^{N})}(x)\simeq ux^{2}$ of the map in the neighborhood of $%
x=0$, there is a map $f^{\ast }(x)$ that satisfies the functional
composition and rescaling equation $f^{\ast }(f^{\ast }(x))=\alpha
^{-1}f^{\ast }(\alpha x)$ for some finite value of $\alpha $ and such that $%
f^{\ast }(x)=ux^{2}+o(x^{4})$. The fixed-point map $f^{\ast }(x)$ possesses
properties common to all superstable attractors of unimodal maps with a
quadratic extremum. Indeed, there is a closed form expression that satisfies
these conditions. This is%
\[
f^{\ast }(x)=x\exp _{q}(u^{q-1}x),
\]%
where $\exp _{q}(x)$ is the $q$-exponential function $\exp _{q}(x)\equiv
\lbrack 1+(1-q)x]^{1/(1-q)}$. The fixed-point map equation is satisfied with 
$\alpha =2^{1/(q-1)}$ and $q=1/2$. The same type of RG solution has been
previously found to exist for the tangent and pitchfork bifurcations of
unimodal maps with general nonlinearity $z>1$ \cite{hu1}, \cite{robledo1}, 
\cite{robledo2}. Use of the map $f^{\ast }(x)$ reproduces the trajectory $%
x_{t}/x_{0}$ and sensibility $\xi _{t}$ shown in Fig. \ref{fig13} (c).

\section{Summary}

We studied the properties of the first few members of the family of
superstable attractors of unimodal maps with quadratic maximum and obtained
a precise understanding of the complex labyrinthine dynamics that develops
as their period $2^{N}$ increases. The study is based on the determination
of the function $t_{f}(x_{0})$, the time of flight for a trajectory with
initial condition $x_{0}$ to reach the attractor or repellor. The novelty
and scope of the study is that the function $t_{f}(x_{0})$ was determined
for \textit{all} initial conditions $x_{0}$ in a partition of the total
phase space $-1\leq x_{0}\leq 1$, and this provides a complete picture for
each attractor-repellor pair. We observed how the fractal features of the
boundaries between the basins of attraction of the positions of the periodic
orbits develop a structure with hierarchy, and how this in turn reflects on
the properties of the trajectories. The set of trajectories produce an
ordered flow towards the attractor or towards the repellor that reflect the
ladder structure of the sub-basins that constitute the mentioned boundaries.
As $2^{N}$ increases there is sensitivity to the final position for almost
all $x_{0}$, and there is a transient exponentially-increasing sensitivity
to initial conditions for almost all $x_{0}$. We observed that transient
chaos is the manifestation of the trajectories' controlled flow out of the
fractal boundaries, that suggest that for large $2^{N}$ the flow becomes an
approximately self-similar sequence of stages. Finally, we looked at the
closing segment of trajectories at which a very fast convergence to the
attractor positions occurs. Here we found `universality class' features, as
the trajectories and sensitivity to initial conditions are replicated by an
RG fixed-point map obtained under functional composition and rescaling. This
map has the same $q$-deformed exponential closed form found to hold also for
the pitchfork and tangent bifurcations of unimodal maps \cite{robledo1}, 
\cite{robledo2}.

The new knowledge gained offers a significant qualitative and quantitative
increment in the understanding of the dynamics at the transition to chaos,
in this case via the period doubling route. The dynamics of approach to the
Feigenbaum attractor and repellor, the supercycle of period $2^{\infty }$,
is, to our knowledge, not known at the level of detail given here for the
first few supercycles. These comprehensive properties of the Feigenbaum
attractor and repellor are presented \cite{moyano2} in tandem with this
work. There is recent renewed interest in the dynamics of low-dimensional
nonlinear maps as access to exact and detailed information for specific
model systems is important in assessing theoretical developments, such as
generalizations of ordinary statistics. There are positive indications that
the multifractal critical attractors present in these maps play such role 
\cite{robledo3}, \cite{robledo4}, \cite{robledo5}.
\\
\\
{Acknowledgments.} D.S. and A.R. are grateful to Hugo Hern\'{a}ndez
Salda\~{n}a for interesting discussions. Partial support by DGAPA-UNAM and
CONACyT (Mexican Agencies) is acknowledged.

\end{document}